\documentclass{lmcs}
\pdfoutput=1

\usepackage{lastpage}
\lmcsdoi{15}{3}{4}
\lmcsheading{}{\pageref{LastPage}}{}{}%
{Jul.~10,~2017}{Jul.~25,~2019}{}

\usepackage{hyperref}
\usepackage{mathrsfs}

\keywords{zero-one laws, limit law, first-order logic, random planar graphs, smooth addable classes}

\title{Logical properties of random graphs \texorpdfstring{\\}{} from small addable classes}
\titlecomment{
    The research reported here was carried out during a visit by the second author to the University of Cambridge in 2011, supported by the ESF Research Networking Programme GAMES\@.
    Eryk Kopczy{\'n}ski was also partially supported by the Polish National Science research grant DEC-2012/07/D/ST6/02435.
}

\author[A. Dawar]{Anuj Dawar\rsuper{a}}
\address{\lsuper{a}Department of Computer Science and Technology, University of Cambridge}
\email{anuj.dawar@cl.cam.ac.uk}

\author[E. Kopczy\'nski]{Eryk Kopczy\'nski\rsuper{b}}
\address{\lsuper{b}Institute of Informatics, University of Warsaw}
\email{erykk@mimuw.edu.pl}

\def\bbN{\mathbb{N}}
\def\bbR{\mathbb{R}}

\def\ra{\rightarrow}

\def\calC{\mathcal{C}}
\def\str{\mathfrak{G}}
\def\struni#1{\str_{U(#1)}} 
\def\Conn{{\rm{Conn}}}
\def\root{{\rm{Root}}}
\def\Var{{\rm{Var}}}
\def\Aut{{\rm{Aut}}}
\def\Vaf{\Var^1}
\def\Vas{\Var^2}

\def\gccnt#1{{Q_{#1}}}

\def\qr{{\rm qr}}
\def\efe{\equiv}

\newcommand{\infL}{L^{\omega}_{\infty\,\omega}}
\newcommand{\valf}{\sigma^1}
\newcommand{\vals}{\sigma^2}
\newcommand{\pfun}{\rightharpoonup}
\newcommand{\power}[1]{\mathscr{P}(#1)} 
\newcommand{\EF}{Ehrenfeucht-Fra\"{\i}ss\'e}
\newcommand{\iso}{\cong}
\newcommand{\disjsum}{\oplus}
\newcommand{\calA}{\mathcal{A}}

\newcommand{\ropen}[1]{[#1)} 
\newcommand{\lopen}[1]{(#1]} 

\begin{document}

\begin{abstract}
  We establish zero-one laws and convergence laws for monadic
  second-order logic (MSO) (and, a fortiori, first-order logic) on a number
  of interesting graph classes.  In particular, we show that MSO obeys
  a zero-one law on the class of connected planar graphs, the class
  of connected graphs of tree-width at most $k$ and the class of
  connected graphs excluding the $k$-clique as a minor.  In each of
  these cases, dropping the connectivity requirement leads to a class
  where the zero-one law fails but a convergence law for MSO still
  holds.
\end{abstract}

\maketitle

\section{Introduction}

The zero-one law for first-order logic~\cite{GKLT69,Fag76} established
that every first-order sentence $\phi$, when evaluated over a random
$n$-element finite structure, has a probability of being true that
converges to either $0$ or $1$ as $n$ goes to infinity.  This prompted
much further investigation into the asymptotic behaviour of classes of
structures definable in logic.  Zero-one laws have been established
for fragments of second-order logic~\cite{KV90}; extensions of
first-order logic such as the infinitary logic $\infL$~\cite{KV92b}
which subsumes various fixed-point logics; and logics with generalized
quantifiers~\cite{DG10}, among many others.

Another widely studied extension of first-order logic is \emph{monadic
  second-order logic} (MSO).  This does not have a zero-one law but
its asymptotic behaviour has been studied on restricted classes of
structures.  For many interesting classes it does admit a zero-one
law, such as on free labelled trees~\cite{zeroonetrees}.  On rooted
labelled trees, MSO does not have a zero-one law, but still admits a
convergence law~\cite{zeroonetrees}.  This means that the probability
of any given sentence $\phi$ being true in an $n$-element structure
does converge to a limit, though that limit is not necessarily $0$ or $1$.
Zero-one and convergence laws for MSO on a number of other classes
are shown in~\cite{Com89}.

In this paper, we are concerned with the asymptotic behaviour of
first-order logic (FO) and MSO on restricted classes of finite
structures, and more specifically, restricted classes of graphs.  We
look particularly at \emph{tame} classes of graphs in the sense
of~\cite{Daw07-mfcs}.  These include planar graphs, graphs of bounded
treewidth and some other proper minor-closed classes.  A flavour of
our results is given by the following examples.  There is a constant
$c \leq 0.036747$ such that the asymptotic probability of any MSO
sentence $\phi$ on the class of planar graphs converges to a real number
in the range $\ropen{0,c} \cup \lopen{1-c,1}$.  On the class of \emph{connected},
planar graphs, MSO admits a zero-one law.  These results can be
strengthened (modulo changing the constant $c$) from planar graphs to
other minor-closed classes of graphs that are \emph{smooth} and
\emph{addable}, which we define formally later.  Here we note that
examples of smooth, addable, minor-closed graph classes include not
just the planar graphs but also the class of graphs of treewidth at
most $k$, for any $k$.

Technically, we rely on the combinatorial results on random planar
graphs proved in~\cite{randompg} and the extensions to minor-closed
classes established in~\cite{randomminor}.  We combine these with
logical techniques from~\cite{Com89} and~\cite{zeroonetrees}.

Since the first submission, it has come to our attention that some of
the results reported here have also been obtained independently
in~\cite{HMNT18}.

\section{Graph Classes}

Let $R_n$ denote a random graph drawn from the uniform distribution on
graphs on the vertex set $[n] = \{1,\ldots,n\}$.  Equivalently, $R_n$
is obtained by putting, for each $i,j$ with $1 \leq i < j \leq n$ an edge between
$i$ and $j$ with probability $1/2$.  If $\calC$ is a class of graphs
closed under isomorphisms, we write $\calC_n$ for the graphs in
$\calC$ on the vertex set $[n]$ and $R_n(\calC)$ to
denote the random graph drawn from the uniform distribution on
$\calC_n$.  Unlike with $R_n$, it is not immediately clear how to
effectively sample from this distribution.  When $\calC$ is the class
of planar graphs, a first partially successful attempt was
in~\cite{markovplanar}, using a Markov chain whose only stable
distribution is the uniform distribution on $\calC_n$.  However, this
is not quite practical as the mixing rate of this Markov chain is
unknown.  Nevertheless, this formulation did enable experimental
validation of some conjectures about random planar graphs.  Finally,
in~\cite{genrandompg}, it is shown that there is a polynomial-time
algorithm that can generate a random planar graph on $[n]$.

An excellent analysis of random planar graphs, resolving some
conjectures of~\cite{markovplanar} was given in~\cite{randompg}.  Some
of this analysis was predicated on a conjecture, termed the ``isolated
vertices conjecture'' to the effect that the number of isolated
vertices in a random planar graph on $[n]$ tends to a limit as $n$
increases.  This conjecture was proved in~\cite{planarsmooth} and
later shown~\cite{randomminor} to be an instance of the
\emph{smoothness} of \emph{addable} minor-closed classes.  Much of the
analysis of random planar graphs in~\cite{randompg} can be extended to
other graph classes that are smooth, addable and
\emph{small}~\cite{randomother}.  We begin by defining these central
notions.

\begin{defi}
  Say that a graph class $\calC$ is \emph{decomposable} if $G \in
  \calC$ if, and only if, every connected component of $G$ is in $\calC$.
\end{defi}

Equivalently, $\calC$ is decomposable if, and only if, $\calC$ is both
closed under disjoint unions and closed under taking connected
components.

\begin{defi}\label{defn:bridge-addable}
  Say that graph class $\calC$ is \emph{bridge-addable} if, for every
  $G \in \calC$, if $u$ and $v$ are vertices in distinct connected
  components of $G$, then the graph obtained by adding the edge
  $\{u,v\}$ to $G$ is also in $\calC$.
\end{defi}

\begin{defi}
  A graph class  $\calC$ is \emph{addable} if it is both decomposable
  and bridge-addable.
\end{defi}

Note that the class of planar graphs is clearly addable.  As noted
in~\cite{randomother}, the following classes of graphs are all
addable: forests; the class of graphs of treewidth at most $k$; the
class of graphs with no cycle of length greater than $k$; the class of
graphs that exclude $K_k$ as a minor.  On the other hand, the class of
graphs embeddable in a torus is not addable.  It is bridge-addable but
not decomposable since it contains $K_5$ but not the graph that is the
disjoint union of two copies of $K_5$.  In general, the class of
graphs of genus at most $k$ for positive $k$ is not addable.

We write $\gccnt{n}(\calC)$ to denote the number of graphs in
$\calC_n$ (i.e.\ graphs in $\calC$ on the vertex set~$[n]$).

\begin{defi}
  A graph class  $\calC$ is \emph{small} if $\gccnt{n}(\calC) \leq d^n n!$ for some $d \in \bbR$. 
\end{defi}

With any class $\calC$, we can associate a \emph{growth constant}
$\gamma_{\calC}$ defined by \[\gamma_{\calC}= \limsup_{n \rightarrow
  \infty} {(\gccnt{n}(\calC)/n!)}^{1/n}.\]  Then, $\calC$ is small just
in case $\gamma_{\calC}$ is finite.

\begin{defiC}[\cite{randomother}]\label{defsmooth}
   A graph class  $\calC$ is \emph{smooth} if  $\frac{\gccnt{n}}{n
     \gccnt{n-1}}$ tends to a finite limit as $n \rightarrow \infty$.
\end{defiC}

It is known (see~\cite{randomother}) that if $\calC$ is smooth, then
the limit of $\frac{\gccnt{n}}{n
     \gccnt{n-1}}$ is the growth constant $\gamma_{\calC}$.   Thus, every smooth class is small.  However,
   there are small graph classes that are not smooth.  When $\calC$ is
   the class of forests, $\gamma_{\calC} = e$.  We can now state the
   two results on smoothness that we need.

\begin{thmC}[\cite{planarsmooth}]
The class of planar graphs is smooth, and the growth constant is $\gamma
\approx 27.22679$ and is given by an explicit analytic expression.
\end{thmC}

\begin{thmC}[\cite{randomminor}]
Each addable proper minor-closed class of graphs is smooth.
\end{thmC}

\section{Logics}

We assume the reader is familiar with the syntax and semantics of
first-order logic (FO) and monadic second-order logic (MSO) interpreted
on finite structures, as defined, for instance in~\cite{Lib04}.  We
give a brief review of definitions, especially where our notation
deviates from the standard.

\subsection{Basics}
We assume we have two disjoint countable sets of symbols $\Vaf$ (the first-order
variables) and $\Vas$ (the second-order variables).  Given a graph
$G=(V,E)$, a \emph{valuation} $\sigma$ on $G$ is a pair
$(\valf,\vals)$ of \emph{partial} functions $\valf: \Vaf \pfun V$ and
$\vals : \Vas \pfun \power{V}$.  For $x \in  \Vaf \cup \Vas$, we use
$\sigma(x)$ to denote $\valf(x)$ or $\vals(x)$ as appropriate.

A \emph{structure} is a pair $\str = (G,\sigma)$ where $G$ is a graph
and $\sigma$ is a valuation on $G$.  If $\str=(G=(V,E), \sigma)$ is a
structure, we sometimes write $G(\str)$, $V(\str)$, $E(\str)$,
$\sigma(\str)$, $x(\str)$ to denote $G, V,
E, \sigma, \sigma(x)$, respectively.  Moreover, we write $\Vaf(\str)$
and $\Vas(\str)$ to denote the subsets of $\Vaf$ and $\Vas$
respectively on which $\valf(\str)$ and $\vals(\str)$ are defined.
For even
greater brevity, when referring to a structure $\str_1$, we may write
$G_1, V_1$, etc.\ for $G(\str_1)$, $V(\str_1)$, etc.\ when no
ambiguity arises.  We write $\sigma[v/x]$ to denote the valuation that
agrees with $\sigma$ at all values other than $x$ and that maps $x$ to
$v$.  We also write $\str[v/x]$ to denote $(G(\str),
\sigma(\str)[v/x])$ and $\str[/x]$ to denote the structure $(G(\str),
\sigma'(\str))$ where the valuation $\sigma'$ is the same as $\sigma$
except that it is undefined at $x$.

Note that, by including the interpretation of variables in our
definition of structure, we can uniformly talk of graphs, coloured
graphs and graphs with distinguished constants as structures, all over a fixed vocabulary.

The formulas of MSO are built-up as usual according to the following
grammar, where $\phi_1$ and $\phi_2$ are formulas, $x,y \in \Vaf$ and
$X,Y \in \Vas$.
\[\phi\ ::=\ x \in X \mid E(x,y) \mid \phi_1 \wedge \phi_2 \mid \phi_1
\vee \phi_2 \mid \neg \phi_1 \mid \forall x \phi \mid \exists x \phi
\mid \forall_2 X \phi \mid \exists_2 X \phi\]
The formulas of first-order logic are those that involve no occurrence
of a variable from $\Vas$.

The definition of satisfaction $\str \models \phi$ is standard.  We
define the \emph{quantifier rank} of a formula $\phi $, denoted $\qr(\phi)$, to be the
maximum depth of nesting of quantifiers in $\phi $ counting both first-
and second-order quantifiers (as in~\cite[Definition 7.4]{Lib04}).

\subsection{Ehrenfeucht-Fra\"{\i}ss\'e games}

We write $\str_1 \efe_m \str_2$ to denote that the two structures
$\str_1$ and $\str_2$ cannot be distinguished by any formula with
quantifier rank at most $m$.  Formally, this is defined by induction
on $m$ as follows.  We say that $\str_1 \efe_0 \str_2$ if, for any
quantifier-free formula $\phi $ such that $\sigma_1$ and $\sigma_2$ are defined on all
the free variables of $\phi $, we have that $\str_1 \models \phi $
if, and only if, $\str_2 \models \phi $.  Inductively, we say that
$\str_1 \efe_{m+1} \str_2$ if all the following conditions are
satisfied.
\begin{itemize}
\item For $x \in \Vaf$, $\forall v_1 \in V_1 \exists v_2 \in V_2 \str_1[v_1/x] \efe_m \str_2[v_2/x]$.

\item For $x \in \Vaf$, $\forall v_2 \in V_2 \exists v_1 \in V_1 \str_1[v_1/x] \efe_m \str_2[v_2/x]$.

\item For $X \in \Vas$, $\forall U_1 \subseteq V_1 \exists U_2 \subseteq V_2 \str_1[U_1/X] \efe_m \str_2[U_2/X]$.

\item For $X \in \Vas$, $\forall U_2 \subseteq V_2 \exists U_1 \subseteq V_1 \str_1[U_1/X] \efe_m \str_2[U_2/X]$.
\end{itemize}

\noindent
As usual, this definition $\str_1 \efe_m \str_2$ can be understood as
a game between two players conventionally called Spoiler and
Duplicator (known as the \emph{$m$-round Ehrenfeucht-Fra\"{\i}ss\'e game}).  The game is played on a ``board''
consisting of two structures $\str_1$ and $\str_2$.   For $m=0$,
Duplicator wins iff $\str_1 \efe_0 \str_2$.  Otherwise, Spoiler
chooses either a variable $x \in \Vaf $ and a vertex $v_i$ in one of
the graphs $G_i$ or a variable $X \in \Vas $ and a set of vertices
$U_i \subseteq V_i$ in the graph $G_i$.  In the first case,
Duplicator responds by choosing an element $v_{3-i} \in G_{3-i}$ and
the play then proceeds for $m-1$ rounds starting with the board
$\str_1[v_1/x]$ and $\str_2[v_2/x]$.  In the second case, Duplicator
responds by choosing a set $U_{3-i} \subseteq V_{3-i}$ and again the play then proceeds for $m-1$ rounds starting with the board
$\str_1[U_1/X]$ and $\str_2[U_2/X]$.  Then, it is clear that
Duplicator has a strategy for winning the $m$ round game on the board
$\str_1$ and $\str_2$ just in case $\str_1 \efe_m \str_2$.  The
connection with MSO comes from the following standard theorem
(see~\cite[Theorem 7.7]{Lib04})
\begin{thm}
The following conditions are equivalent:
\begin{itemize}
\item $\str_1 \efe_m \str_2$
\item $\str_1$ and $\str_2$ satisfy exactly the same MSO formulas of quantifier rank at most $m$.
\end{itemize}
\end{thm}

\noindent
Finally, we note that, if we have only a fixed finite set of
variables, then the equivalence relation $\efe_m$ is of finite index
for each $m$.  That is, fix a finite set $\Xi \subseteq \Var$ and let
$\mathcal{S}[\Xi]$ denote the class of structures $\str$ such that
$\Vaf(\str), \Vas(\str) \subseteq \Xi$.
\begin{prop}
  The relation $\efe_m$ restricted to $\mathcal{S}[\Xi]$ has finite
  index.
\end{prop}
The proof is a proof by induction that there are, up to equivalence,
only finitely many formulas of MSO of quantifier rank $m$ with free
variables among $\Xi$ (see~\cite[Prop.~7.5]{Lib04}).  We write
$t_m(\Xi)$ to denote the index of $\efe_m$ restricted to
$\mathcal{S}[\Xi]$.  Note that the value of $t_m(\Xi)$ is completely
determined by the number of elements in $\Xi \cap \Vaf$ and $\Xi \cap
\Vas$.

\section{Adding graphs}

For structures $\str_1$ and $\str_2$, the disjoint union $\str_1
\disjsum \str_2$ is defined if $\Vaf_1 \cap \Vaf_2 = \emptyset$.
In this case, the set of vertices of $\str_3 = \str_1 \disjsum \str_2$ is the
disjoint union of $V_1$ and $V_2$, $E_3 = E_1 \cup E_2$,
$\sigma^1_3(x) = \sigma_i^1(x)$ for $i \in \{1,2\}$ whenever $x \in \Vaf_i$
and $\sigma^2_3(X) = \sigma_1^2(X) \cup \sigma_2^2(X)$.

It is well-known that the equivalence relation $\efe_m$ is a
congruence with respect to the disjoint union of structures.
Moreover, for every $m$ and finite set $\Xi \subseteq \Vaf \cup \Vas$ there is a threshold $r_m(\Xi)$ such that for any
$p,q > r_m(\Xi)$, the disjoint union of $p$ copies of a structure is
$\equiv_m$-equivalent to the disjoint union of $q$ copies.  We
formally state this for later use.
\begin{lem}\label{lem:disjoint-sum}
Fix a positive integer $m$ and a finite set $\Xi \subseteq \Vaf \cup
\Vas$.  There is a positive integer $r_m(\Xi)$ such that
  if $\{ \str_i \mid i \in I\}$ is a collection of structures such
  that $\str_i \equiv \str_j$ for all $i,j\in I$ and $P, Q \subseteq
  I$ are sets such that either $|P| = |Q|$ or $|P|, |Q| \geq r_m(\Xi)$ then
\[ \bigoplus_{i \in P} \str_i \efe_m \bigoplus_{j \in Q} \str_j. \]
\end{lem}

Indeed,
similar facts can be established for many operations other than
disjoint union (see, for example~\cite{Makowsky04}).
Here, we are also
interested in a particular operation of taking the disjoint union of
two structures while adding an edge between distinguished vertices (as
in the definition of bridge addable graphs:
Definition~\ref{defn:bridge-addable}).  We give the formal definition
below and prove the properties we need.

We say that a structure $\str$ is \emph{rooted} if $\root \in \Vaf(\str)$.
Let $\str_1$ and $\str_2$ be two rooted structures such that $\Vaf_1
\cap \Vaf_2 = \{\root\}$ and $V_1 \cap V_2 = \emptyset$.  We write
$\str_1 + \str_2$ to denote the structure $\str$ such that:

\begin{itemize}
\item $V = V_1 \cup V_2$, $E = E_1 \cup E_2 \cup \{\sigma_1(\root), \sigma_2(\root)\}$
\item $\sigma(X) = \sigma_1(X) \cup \sigma_2(X)$ for $X \in \Vas_1 \cap \Vas_2$
\item $\sigma(X) = \sigma_i(X)$ for $X \in \Vas_i - \Vas_{3-i}$
\item $\sigma(\root) = \sigma_1(\root)$
\item $\sigma(x) = \sigma_i(x)$ for $x \in \Vaf_i$ and $x \neq \root$
\end{itemize}
That is to say that $\str_1 + \str_2$ is obtained by taking the
disjoint union of the structures $\str_1$ and $\str_2$ seen as
coloured graphs, putting an edge between the two roots, and making the
root of $\str_1$ the root of the combined structure.  This asymmetry
in the choice of root means that, in general  $\str_1 + \str_2 \neq
\str_2 + \str_1$.  However, it is still the case that $(\str_1 +
\str_2) + \str_3 = (\str_1 + \str_3) + \str_2$.

A simple argument using Ehrenfeucht-Fra\"{\i}ss\'e games establishes
the following proposition.

\begin{prop}\label{prop:congruence}
If $\str_1 \efe_m \str_2$, then $\str_0 + \str_1 \efe_m \str_0 + \str_2$.
\end{prop}
\begin{proof}
Suppose Duplicator has a winning strategy in the $m$-move
Ehrenfeucht-Fra\"{\i}ss\'e game played on $\str_1$ and $\str_2$.  We
show that she also has a winning strategy in the $m$-move game
on $\str_0 + \str_1$ and $\str_0 + \str_2$.  The strategy is described
as follows.
\begin{itemize}
\item if Spoiler chooses $w \in V_0$, then Duplicator responds with the same $w$;
\item if Spoiler chooses $w \in V_i$ for $i\in \{1,2\}$, then
  Duplicator responds with $w' \in V_{3-i}$ given by her winning
  strategy in the game on  $\str_1$ and $\str_2$; and
\item if Spoiler chooses $W \subseteq V_0 \cup V_i$ for  $i\in
  \{1,2\}$, then Duplicator responds with $(W \cap V_0) \cup W_2$,
  where $W_2$ is her response to $W \cap V_i$ in the game on  $\str_1$ and $\str_2$.
\end{itemize}
It is easy to show that this does indeed describe a winning strategy.
\end{proof}

If $\str_2$ is a structure with $\Vaf_2= \{\root\}$ and $c$ is a
positive integer, we write  $\str_1+c\str_2$ to denote the structure
obtained by adding $\str_2$ to $\str_1$ $c$ times.  More formally,
this is defined by induction on $c$:
\begin{itemize}
\item $\str_1+0\str_2 = \str_1$,
\item $\str_1+(c+1)\str_2 = (\str_1+c\str_2) + \str_2$
\end{itemize}

\noindent
Analogously to Lemma~\ref{lem:disjoint-sum}, we have the following
proposition.
\begin{prop}\label{prop:copies}
  Let $\Xi \subseteq \Var$ be a finite set.  For each $m$, there is
  number $q_m(\Xi)$ such that for any $\str \in\mathcal{S}[\Xi]$ and
  any structure $\str_0$ we have
  \[ \str_0+p\str \efe_m \str_0+q \str \]
  whenever $p,q \geq q_m(\Xi)$.
\end{prop}
\begin{proof}
  We define the value of $q_m(\Xi)$ by induction on $m$,
  simultaneously for all finite sets $\Xi \subseteq \Var$.
  \begin{itemize}
  \item $q_0(\Xi) = 0$ for all $\Xi$; and
  \item $q_{m+1}(\Xi) = \max(q_m(\Xi \cup \{x\})+1,\ t_m(\Xi \cup \{X\})\cdot q_m(\Xi \cup \{X\}) +
    m)$,
  \end{itemize}
where $x \in \Vaf,\ X \in \Vas$ are any variables not in $\Xi$.  Recall that $t_m(\Xi)$
denotes the index of the relation $\efe_m$ in $\mathcal{S}[\Xi]$.

Now, we argue by induction on $m$ that if $p,q \geq q_m(\Xi)$ then
$\str_0+p\str \efe_m \str_0+q \str$.   In the following, we use the
sets $[p]$ and $[q]$ to index the copies of $\str$.  Thus, we may
refer to the graph $G_i$  for $i \in [p]$.
The base case, $m=0$ is immediate.  Suppose now that the claim is true for some $m$ and all finite
$\Xi$, and let  $\str_A = \str_0+p\str$ and $\str_B = \str_0+q \str$
for some $p,q \geq q_{m+1}(\Xi)$.  We consider a number of cases
corresponding to different moves that Spoiler might make in the \EF\
game.
\begin{itemize}
\item Suppose $x \in \Vaf$ and $v \in V_0$.  Then,
\[
\begin{array}{lcl}
(\str_0+ p\str)[v/x] & = & \str_0[v/x] + p\str \\ & \efe_m &
\str_0[v/x]+q\str  \\ & = & (\str_0+q\str)[v/x],
\end{array}
\]
where the central $\efe_m$ is true by induction hypothesis, since
$p,q \geq q_{m+1}(\Xi) > q_m(\Xi \cup \{x\})$.
\item Suppose $x \in \Vaf$ and $v \in V_i$  for $i \in [p]$.  Then
\[
\begin{array}{lcl}
(\str_0+ p\str)[v/x] & \iso & (\str_0 + \str)[v/x] + (p-1)\str
\\ & \efe_m &
 (\str_0 + \str)[v/x] + (q-1)\str \\ & \iso &  (\str_0+q\str)[v/x],
\end{array}
\]
where the central $\efe_m$ is again  true by induction hypothesis, since
$p-1,q-1 \geq q_m(\Xi \cup \{x\})$.  The situation where  $v \in V_i$
for $i \in [q]$ is entirely symmetric.
\item Suppose now that $X \in \Vas$ and $U \subseteq V_A$.  We
  partition $[p]$ into sets $P_1,\ldots,P_t$ where $t \leq  t_m(\Xi \cup
  \{X\})$ such that $i$ and $j$ are in the same part if, and only if,
  $\str_i[U \cap V_i/X] \efe_m \str_j[U \cap V_j/X]$.  Since $p,q \geq
  q_{m+1}(\Xi) \geq t_m(\Xi \cup \{X\})\cdot q_m(\Xi \cup \{X\})$ we
  can partition $[q]$ into parts $Q_1,\ldots,Q_t$ such that for all
  $k$, either $P_k$ and $Q_k$ have the same size, or they both have
  more than $q_m(\Xi \cup \{X\})$ elements.  For each $l$ choose an
  $i_l \in P_l$ and define, for $j \in [q]$, $U'_j$ to be $U \cap
  V_{i_l}$ whenever $j \in Q_l$.  Let $U' = \bigcup_{j \in[q]}U'_j$.  Then, by the induction hypothesis
  and Proposition~\ref{prop:congruence}, it follows that $\str_A[U/X]
  \efe_m \str_B[U/X]$.  Again, the case when $U \subseteq V_B$ is
  entirely symmetric.
  \qedhere
\end{itemize}

\end{proof}

\subsection{Universal Connected Structure}\label{sec:universal}

There is a sentence of MSO (with quantifier rank 5) which is true in a
structure $\str$ if, and only if, $\str$ is connected.  It follows
that if $m \geq 5$ then each $\efe_m$ class of structures either
contains only connected structures or disconnected ones.

For the rest
of this subsection, we fix a value $m$ with $m \geq 5$.  Also, let
$\calC$ be a class of graphs closed under the operation $+$.  Let
$\str_1,\ldots,\str_t$ (with $t < t_m(\emptyset)$) be a set of
representatives from $\calC$ of all $\efe_m$ classes of connected graphs which have
elements in $\calC$.  Let
$\str_R$ denote the rooted structure with one element, no edges, and
interpreting no variables other than Root.  We define the
\emph{$m$-universal connected rooted graph in $\calC$} to be the
structure  $\struni{m} = \str_R + \sum_{1 \leq i \leq t} q_k(\emptyset)\str_i$.
That is,  $\struni{m}$ is obtained by adding $q_k(\emptyset)$ copies
of a representative of each $\efe_m$ class of graphs to $\str_R$.
Changing the order in which these graphs are added does not change the
isomorphism type of $\struni{m}$ and changing the choice of
representatives $\str_1,\ldots,\str_t$ does not affect the $\efe_m$
class of  $\struni{m}$.

\begin{defi}
  Say that a rooted structure $\str$ \emph{appears} in a graph $\str_1$
  if there is an induced substructure $\str_2$ of $\str_1$ and a vertex $r \in
  V_2$ such that:
  \begin{itemize}
  \item $\str$ is isomorphic to $\str_2[r/\root]$; and
  \item there is only one edge between $V_2$ and $V_1\setminus V_2$
    and this edge is incident on $r$.
  \end{itemize}
\end{defi}

\begin{prop}\label{appcorol}
  If the universal structure $\struni{m}$ appears in a connected graph $\str$, then $\str
  \efe_m \struni{m}[/\root]$.
\end{prop}
\begin{proof}
  By the definition of appearance, there is a vertex $r$ in $\str$
  such that $\str[r/\root]$ is isomorphic to $\struni{m}+\str'$ for
  some $\str'$.  Let $i$ be such that $\str_i$ with $\str_i \efe_m
  \str'$ is the representative of the $\efe_m$ equivalence class of
  $\str'$ in the definition of $\struni{m}$.  Then, by
  Prop.~\ref{prop:congruence}, $\str[r/\root] \efe_m \struni{m} +
  \str_i$.  Since, $\struni{m}$ contains more than $q_m(\emptyset)$
  copies of $\str_i$, Prop,~\ref{prop:copies} gives us that
  $\struni{m} \efe_m \struni{m} + \str_i$ and the result follows.
\end{proof}

\section{MSO zero-one law for random connected graphs}\label{sec:connected}

Let $\calC$ be a class of graphs.  Recall that $\calC_n$ is the class
of graphs in $\calC$ on the vertex set $[n]$ and
$R_n(\calC)$ denotes the random graph drawn from the uniform distribution on
$\calC_n$.  The following is a direct consequence of Theorem~1.1
in~\cite{ChapuyP16}.
\begin{thm}\label{usuallyconnected}
Let $\calC$ be a small addable class of graphs. Then $R_n(\calC)$ is
connected with probability at least $1/\sqrt{e} - o(1)$.
\end{thm}
Indeed, Theorem~1.1 of~\cite{ChapuyP16} establishes this more
generally for \emph{bridge addable} classes.  The bound $1/\sqrt{e}$
is tight in that this is the limiting probability for connectedness
among forests.

We also rely on the following result
which is obtained as a consequence of Theorem~5.1
in~\cite{randomother}.

\begin{thm}\label{alwaysappear}
Let $\calC$ be a small addable class of graphs, and $\str$ be a rooted
graph in $\calC$. Then the probability that $\str$ appears in $R_n(\calC)$ tends
to 1 as $n \ra \infty$.
\end{thm}

In fact, Theorem~5.1 of~\cite{randomother} establishes the stronger
result that the number of appearances of $\str$ in $R_n(\calC)$ grows
linearly with $n$ (more precisely, there is an $\alpha > 0 $ such that the
 probability that $\str$ appears $\alpha n$ times tends to 1).

Together, these enable us to establish the zero-one law for the class
of connected graphs in any small addable class $\calC$.  In the
following we write \Conn\ for the class of \emph{connected} graphs.
\begin{thm}\label{zerooneconn}
Let $\phi$ be a sentence of MSO, and $\calC$ a small addable class of
graphs.  Let $p_n$ denote the probability that $R_n(\calC \cap \Conn)$
satisfies $\phi$. Then $\lim_{n \rightarrow \infty} p_n$ is defined
and equal to either $0$ or $1$.
\end{thm}
\begin{proof}
  Let $m$ be the quantifier rank of $\phi$.  By Theorem~\ref{alwaysappear}
  we know that the probability that $\struni{m}$
  appears in $R_n(\calC)$ tends to 1.  Moreover, since the probability
  that $R_n(\calC)$ is connected is non-zero (by
  Theorem~\ref{usuallyconnected}) it follows that the probability that
  $\struni{m}$, the $m$-universal connected rooted graph in $\calC$, appears in $R_n(\calC \cap \Conn)$ also tends to $1$.
  Thus, by Proposition~\ref{appcorol}, with probability tending to
  $1$, we have $R_n(\calC \cap \Conn) \efe_m \struni{m}$.  Hence, if
  $\struni{m} \models \phi$, then $p_n$ tends to $1$, otherwise $p_n$
  tends to $0$.
\end{proof}

As an immediate consequence, we have a zero-one law for MSO for a
number of interesting classes of graphs.

\begin{cor}
  MSO admits a zero-one law on each of the following classes of
  graphs.
  \begin{itemize}
  \item The class of connected planar graphs.
  \item For each $k$, the class of connected graphs of tree-width at
    most $k$.
  \item For each $k > 2$, the class of connected graphs excluding $K_k$ as
    a minor.
  \end{itemize}
\end{cor}

\section{MSO limit law for random graphs}

We are now ready to establish our general result on the existence of
a limit law (also known as a convergence law) for MSO on smooth addable
classes of graphs.  Note that, while Theorem~\ref{zerooneconn} was
stated for small addable classes, from now on we will restrict
ourselves further to \emph{smooth} classes.  Recall that every smooth
class is also small.

To establish the limit law we need two specific results
from~\cite{randomother}.  The first is a direct consequence of
Theorem~6.4 in that paper.
\begin{thm}\label{giantcomp}
Let $\calC$ be a smooth addable class of graphs.
For any $\epsilon > 0$ there exist a constant $g(\epsilon)$
such that, for sufficiently large values of $n$, with probability at
least $1-\epsilon$, $R_n(\calC)$ has a
connected component which contains at least $n - g(\epsilon)$ vertices.
\end{thm}
In general, we refer to the \emph{giant component} of $R_n(\calC)$.

\medskip

The second result we need is a consequence of  Theorem 9.2
of~\cite{randomother}.
\begin{thm}\label{poisson}
Let $\calC$ be a smooth addable class of graphs, $H$ be a
connected graph, and $k \in \bbN$. Then there is a number $p_k(H)$
such that the probability that $R_n(\calC)$ has exactly $k$ components
isomorphic to $H$ tends to $p_k(H)$. Moreover, these events are
asymptotically independent for non-isomorphic graphs $H$.
\end{thm}
Actually, Theorem 9.2 of~\cite{randomother} does not state that the
events are asymptotically independent, but it is easily seen to be the case.  Indeed,
what is stated there is that the distribution of the number of
components  isomorphic to $H$ tends to Poisson distribution as $n \ra \infty$,
with parameter $\lambda = 1 / ({\gamma(\calC)}^{|H|} |\Aut(H)|)$, where
$|\Aut(H)|$ is the number of automorphisms of $H$.

As an example,  for large values of
$n$, the random planar graph on $n$ vertices has on average $1/\gamma(\calC) \approx 0.03673$
isolated vertices, $1/2{\gamma(\calC)}^2 < 0.0007$ pairs of vertices
with an edge between them (and no other edges incident on them), less than $0.00004$ isolated connected subgraphs with
3 vertices, and so on.  Summing over all of these, we can show that the random planar
graph has $\lambda \approx 0.037439$ connected components other than the
giant component on average.  From the fact that the distribution is
Poisson we get that the probability that the graph
is not connected is $1-e^{-\lambda} \approx 0.036746$.

We are interested in the frequency of occurrence of connected
components, not just up to isomorphism, but up to $\efe_m$ for suitable
values of $m$.  Specifically, we are interested in whether the number
of components from a fixed $\efe_m$ equivalence class is greater than
the threshold $r_m(\emptyset)$ from Lemma~\ref{lem:disjoint-sum} and
the exact number if it is not.  In the following result, we use
Theorems~\ref{giantcomp} and~\ref{poisson} to show that the relevant
probabilities converge.
\begin{thm}\label{thm:mso-poisson}
  Let  $\calC$ be a smooth addable class of graphs and $m \in
  \bbN$. If $\calA$ is an $\efe_m$-equivalence class of connected graphs
    then for each $k \leq r_m(\emptyset)$
    there is a $p_k(\calA) \in \bbR$ such that the probability that
    $R_n(\calC)$ has exactly $k$ components from $\calA$ tends to
    $p_k(\calA)$ as $n$ goes to infinity.
\end{thm}
\begin{proof}
Let $p_n$ denote the probability that $R_n(\calC)$ contains exactly
$k$ components in $\calA$.  We
show that for any $\epsilon > 0$ there is a $p$ such that $|p_n -p| <
\epsilon$ for large enough $n$.  Thus, the sequence ${(p_n)}_{n \in \bbN}$ is a Cauchy
sequence and so converges to a limit.

Let $g = g(\epsilon/3)$ be the value given by Theorem~\ref{giantcomp}
such that, for sufficiently large $n$, with probability at least $1-
\epsilon/3$, $R_n(\calC)$ has a
connected component which contains at least $n - g$ vertices.
Moreover, let $n_0 \in \bbN$ be such that for all $n > n_0$, this
probability is indeed at least $1-\epsilon/3$.

By exactly the same argument as in the proof of
Theorem~\ref{zerooneconn}, we can show that the conditional
probability, given that  $R_n(\calC)$ has a
connected component which contains at least $n - g$ vertices, that
this giant component is $\efe_m$-equivalent to $\struni{m}$, the $m$-universal connected rooted graph in $\calC$, tends to
$1$ as $n$ goes to infinity.  Thus, we can fix a value $n_1 \in \bbN$
such that this probability is at least $1-\epsilon/3$ for all $n >
n_1$.

Let $H_1, \ldots, H_M$ enumerate (up to isomorphism) all graphs in
$\calA$ with at most $g$ vertices.  Let $K$ denote the collection of
all functions $f: [M] \ra \bbN$ such that $\sum_{i \in [M]}f(i) = k$
and note that this is a finite set.  Let $p = \sum_{f \in K}\prod_{i
  \in [M]} p_{f(i)}(H_i)$, where $p_{f(i)}(H_i)$ is the limiting
probability, given by Theorem~\ref{poisson}, that $R_n(\calC)$ contains
exactly $f(i)$ components isomorphic to $H_i$.  If $p'_n$ denotes the
probability that $R_n(\calC)$ contains exactly $k$
components that are isomorphic to one of $H_1, \ldots, H_M$, then
clearly the sequence ${(p'_n)}_{n \in \bbN}$ tends to the limit $p$.
Let $n_2 \in \bbN$ be such that $|p'_n - p| < \epsilon/3$ for all $n
> n_2$.

First, consider the case that $\struni{m} \not\in \calA$, i.e.\
$\calA$ is an $\efe_m$-equivalence class distinct from that of
$\struni{m}$.  In this case, our aim is to show that for all sufficiently large $n$, in particular
for all $n > \max(n_0,n_1,n_2)$, we have $p - \epsilon < p_n < p + \epsilon$,
establishing the result.  We return to the case where  $\struni{m} \in
\calA$ later.

Fix $n$ with $n > \max(n_0,n_1,n_2)$ and let $p_0$ denote the probability that $R_n(\calC)$ contains \emph{no}
component from $\calA$ except those that are isomorphic to one of
$H_1, \ldots, H_M$.  Then clearly $p_n > p'_n\cdot p_0$ since the
left-hand side denotes the probability that there are exactly $k$
components from $\calA$ and the right-hand side gives the probability
of obtaining exactly $k$ components from $\calA$ in a particular way,
i.e.\ all from among $H_1, \ldots, H_M$.  Moreover $p_0 >
{(1-\epsilon/3)}^2$ since  if there is a giant component with
$n-g$ elements and it is $\efe_m$-equivalent to $\struni{m}$ then
there cannot be a component with more than $g$ vertices from $\calA$.
We then have
\[
\begin{array}{lcl}
p_n & > & p'_n p_0 \\
&  > & {p'_n(1-\epsilon/3)}^2 \\
& > & {p(1-\epsilon/3)}^3 \\
& > & p - \epsilon,
\end{array}
\]
where the second line follows by substituting the lower bound for
$p_0$ and the third line follows from the fact that $|p'_n - p| <
\epsilon/3$.

For the other direction, note that
\[ p_n < p_n''p_0 + (1-p_n')(1-p_0),\]
where the first term on the right is as above and the second term is
an upper bound on the probability that $k$ components from $\calA$ are
obtained in some way other than by having $k$ components with $g$
vertices or fewer and $0$ components with more than $g$ vertices.
We then have
\[
\begin{array}{lcl}
p_n & < & p'_n p_0 + (1-p'_n)(1-p_0)\\
&  < & p'_n+ (1-p_0) \\
& < & p'_n + (1-{(1-\epsilon/3)}^2) \\
& < & p + \epsilon/3 + (1-{(1-\epsilon/3)}^2)\\
& < & p + \epsilon,
\end{array}
\]
where the second line is obtained by substituting the upper bound of
$1$ for $p_0$ and $1-p'_n$, the third line by substituting the lower
bound of ${(1-\epsilon/3)}^2$ for $p_0$ and the fourth by  the fact that $|p'_n - p| <
\epsilon/3$.

For the case that $\struni{m} \in \calA$, an entirely analogous
argument can be used to show that $|p_n - p| < \epsilon$ where $p$ is
the limit of the sequence $p''_n$ of probabilities that  $R_n(\calC)$ contains exactly $k-1$
components that are isomorphic to one of $H_1, \ldots, H_M$.
\end{proof}

Fix a small addable class of graphs $\calC$ and $m \in \bbN$.  Let $r
= r_m(\emptyset)$ and $t = t_m(\emptyset)$.  Suppose
$\calA_1,\ldots,\calA_t$ enumerates all the $\efe_m$ classes of graphs
in $\calC$.  We call an
\emph{$m$-profile} a function $f: [t] \ra \{0,\ldots,r\}$.  We say
that a graph $G$ \emph{matches} the $m$-profile $f$ if the following conditions
hold:
\begin{enumerate}
\item for each $i \in [t]$, if $f(i) < r$ then $G$ has exactly $f(i)$
  connected components which are in $\calA_i$; and
\item for each $i \in [t]$, if $f(i) = r$ then $G$ has at least $r$
  distinct connected components which are in $\calA_i$.
\end{enumerate}

The following lemmas are now immediate from our previous
constructions.
\begin{lem}\label{lem:profile-converge}
  If $\calC$ is a smooth addable class of graphs then for any $m$ and
  any $m$-profile $f$, the probability
  that $R_n(\calC)$ matches $f$ converges to a value $p_f$ as $n$ goes
  to infinity.
\end{lem}
\begin{proof}
Define $p_f$ to be
\[p_f = \left( \prod_{i : f(i) < r} p_{f(i)}(\calA_i) \right)
\left(\prod_{i : f(i) = r} \sum_{k \geq r} p_k(\calA_i) \right).\]
The result than follows by the asymptotic independence asserted in
Theorem~\ref{poisson} above.
\end{proof}

Note that every graph matches some profile $f$ and these events are
mutually exclusive for distinct $f$.  Thus, the sum of
$p_f$ over all profiles $f$ must be $1$.  The reason for considering
profiles is, of course, that they completely determine the $\efe_m$
class of a graph.
\begin{lem}\label{lem:profile-equivalent}
  If $G_1$ and $G_2$ are graphs that both match the same $m$-profile $f$,
  then $G_1 \efe_m G_2$.
\end{lem}
\begin{proof}
  This is immediate from Lemma~\ref{lem:disjoint-sum}.
\end{proof}

We can now establish the main convergence law for MSO\@.
\begin{thm}\label{thm:limit}
  If $\calC$ is a smooth, addable class of graphs and $\phi$ is a
  sentence of MSO, then the probability that $R_n(\calC)$ satisfies
  $\phi$ tends to a limit as $n$ goes to infinity.
\end{thm}
\begin{proof}
  Let $m$ be the quantifier rank of $\phi$.  By
  Lemma~\ref{lem:profile-equivalent}, if $f$ is an $m$-profile then
  either all graphs matching $f$ satisfy $\phi$ or none do.  Let us
  say that $f$ implies $\phi$ if the former case holds.  Then taking
  $p$ to be the sum of $p_f$ (as in Lemma~\ref{lem:profile-converge})
  over all $f$ that imply $\phi$,
  we see that the probability that  $R_n(\calC)$ satisfies
  $\phi$ tends to $p$.
\end{proof}

We can say somewhat more about the possible values $p$ to which the
probability that $\phi$ holds in $R_n(\calC)$ may converge.  Note that
the property of being connected is definable by a sentence of MSO and
thus the probability that $R_n(\calC)$ is connected converges to a
limit.  By Theorem~\ref{usuallyconnected}, this value is at least
$1/\sqrt{e}$ and therefore greater than $1/2$.
  This, together with the result below tells us that the
limiting probabilities of MSO sentences on smooth, addable classes
cluster near $0$ and $1$.

\begin{thm}
  Let $\calC$ be a smooth, addable class and let $c$ be the
  limiting probability that $R_n(\calC)$ is not connected.  Then, for any
  MSO sentence $\phi$, the probability that $R_n(\calC)$ satisfies
  $\phi$ converges to a value $p$ such that either $p \leq c$ or $p
  \geq 1-c$.
\end{thm}
\begin{proof}
  Let $m$ be the quantifier rank of $\phi$ and let
  $\struni{m}$ be the universal connected structure defined in
  Sec.~\ref{sec:universal}.  If $\struni{m} \models \phi$ then $\phi$
  is true in  $R_n(\calC \cap \Conn)$ with probability tending to
  $1$.  Thus the probability that  $R_n(\calC)$ satisfies $\phi$ tends
  to at least $1-c$.  On the other hand, if $\struni{m} \not\models
  \phi$, then  $\phi$
  is false in  $R_n(\calC \cap \Conn)$ with probability tending to
  $1$, and so the probability that  $R_n(\calC)$ satisfies $\phi$ tends
  to at most $c$.
\end{proof}

For many interesting classes of graphs, the value of $c$ is quite
small.  As noted above, for planar graphs the value of $c$ is about
$0.036746$, giving us the result mentioned in the introduction.

\section{Future work}

We have shown that the zero-one law holds for random connected graphs from
smooth addable classes of graphs, for formulas of first order and
monadic second-order logic.  Moreover, a limit law holds for random graphs
of such classes which do not have to be connected.  This includes many
of the tame classes of graphs that have been studied in finite model
theory in recent years.  Still, there are other classes one could
explore.  Most interesting would be proper minor-closed classes which
are not addable, such as the graphs embeddable in a torus or, more
generally, the class of graphs of genus at most $k$ for a fixed value
of $k$.

Another general direction would be to explore logics beyond
first-order logic (other than MSO) such as fixed-point logics.  These
are known to admit a zero-one law over the class of all graphs.
However, their study is based on equivalence relations $\equiv^k$
stratified by the number of variables rather than the quantifier
rank.  These equivalence relations do not have finite index and that
makes many of the methods we have used here infeasible to use.

\bibliographystyle{alpha}
\bibliography{zeroone}{}

\end{document}